\documentclass[aps,pra,twocolumn]{revtex4}
\usepackage[usenames,dvipsnames]{color}
\usepackage{graphicx}
\usepackage{amssymb}
\usepackage{subfigure}
\usepackage{stackrel}
\usepackage{amsmath}
\usepackage{amsfonts}
\usepackage{bm}
\usepackage[normalem]{ulem}
\usepackage{hyperref}
\newcommand{\vect}[1]{\ensuremath{\bm{{#1}}}}
\newcommand{\ket}[1]{\ensuremath{\left|{#1}\right\rangle}}
\newcommand{\bra}[1]{\ensuremath{\left\langle{#1}\right |}}

\newcommand{\be}{\begin{equation}}
\newcommand{\ee}{\end{equation}}
\newcommand{\ben}{\begin{eqnarray}}
\newcommand{\een}{\end{eqnarray}}

\usepackage{graphicx}
\usepackage{color} 
\usepackage{bm}   

\begin{document}

\title{Emergent Dynamics from Entangled Mixed States}

\author{A. Vald\'es-Hern\'andez$^1$, C. G. Maglione$^{2}$, A. P. Majtey$^{2,3}$, and A. R. Plastino$^4$}
\address{$^1$Instituto de F\'{\i}sica, Universidad Nacional Aut\'{o}noma de M\'{e}xico,
Apartado Postal 20-364, Ciudad de M\'{e}xico, Mexico.}
\address{$^2$Universidad Nacional de C\'ordoba, Facultad de Matem\'atica, Astronom\'{\i}a, F\'{\i}sica y Computaci\'on, Av. Medina Allende s/n, Ciudad Universitaria, X5000HUA C\'ordoba, Argentina.}
\address{$^3$ Instituto de F\'iÂ­sica Enrique Gaviola (IFEG), Consejo Nacional de Investigaciones Cient\'{i}ficas y T\'ecnicas de la Rep\'ublica Argentina (CONICET), C\'ordoba, Argentina.}
\address{$^4$CeBio y  Departamento de Ciencias B\'asicas, Universidad Nacional del Noroeste de la Prov. de Buenos Aires, UNNOBA, CONICET, Roque Saenz-Pe\~na 456, Junin, Argentina.\\}
\vspace{10pt}

\begin{abstract}
Entanglement is at the core of quantum physics, playing a central
role in  quantum phenomena involving composite systems.
According to the \emph{timeless picture of quantum dynamics},
entanglement may also be essential for understanding the very
origins of dynamical evolution and the flow of time. Within this
point of view, the Universe is regarded as a bipartite entity comprising
a clock $C$ and a system $R$ (or ``rest of the Universe") jointly
described by a global stationary state, and the dynamical evolution
of $R$ is construed as an emergent phenomena arising from the
entanglement between $C$ and $R$. In spite
of substantial recent efforts, many aspects of this approach
remain unexplored, particularly those involving mixed states.
In the present contribution we investigate the
timeless picture of quantum dynamics for mixed states of the
clock-system composite, focusing on quantitative relations linking
the clock-system entanglement with the emerging dynamical evolution
experienced by the system.

\end{abstract}
\maketitle

\section{Introduction}

  One of the goals of Science is to formulate the most economical description
  possible of natural phenomena. Guided by this desire for conceptual economy,
  scientists try to develop theories having the least possible number of basic
  assumptions or primitive elements. In this regard, research into the
  phenomenon of quantum entanglement has led to remarkable insights.
  For instance, the study of entanglement  clarified
  the origin of the states describing systems in thermal equilibrium with
  a heat bath, without the need to invoke the micro-canonical distribution
  for the system-bath composite \cite{GLTZ2006,DZ2016}. More radically, research work revolving around
  quantum entanglement also  provided a plausible explanation of the origins
  of dynamical evolution and the flow of time. The concomitant arguments,
  according to which time and dynamics
  are emergent phenomena arising from quantum correlations, were first articulated by
  Page and Wootters (PW) \cite{PW1983,W1984}, although related ideas had been previously advanced
  in the context of the quantum theory of gravity \cite{DeW1967, HH1983}.

  Within the PW \textit{timeless} picture of quantum mechanics
  \cite{PW1983,W1984}, the whole Universe $U$ is assumed to be in a global
  stationary state, which is an eigenstate of the total Hamiltonian with zero energy
  eigenvalue. Dynamical evolution arises from this static state as a
  result of the quantum entanglement between the degree of freedom of an appropriate subsystem $C$, called the \textit{clock}, and the \textit{rest of the Universe} $R$. According to this idea, time and dynamics are emergent features of the Universe rooted in the entanglement between two subsystems, $R$ and $C$.
  The Schr\"odinger time-independent equation describing the global stationary state of the $R+C$
  composite is reminiscent of the celebrated Wheeler-DeWitt equation in quantum cosmology,
  describing a stationary state with zero eigenvalue for the wave function of the entire (closed) Universe \cite{DeW1967, HH1983}.

  The PW timeless approach to quantum mechanics has been elaborated and extended
  in various directions, from both the theoretical and the experimental points of view  \cite{R1990,GPPT,AI2008,V2014,MV2017,GLM2015,BRGC2016,BR2018,LM2017,MS2019,
  DR2019,DMR2019,MPA2013,MBGGMG2014,MSVW2015,PRBGBRL2019,S2018,MVMP2019}.
  Healthy controversy \cite{AI2008,MV2017} has invigorated research into the PW
  proposal, stimulating the exploration of its possibilities. The timeless picture was
  criticized by Albrecht and Iglesias \cite{AI2008}, who pointed out apparent ambiguities
  concerning non-equivalent choices for the clock subsystem. Subsequent counter arguments
  by Marletto and Vedral \cite{MV2017} showed that these ambiguities do not arise, if
  one takes carefully into account the properties needed by a subsystem to be
  acceptable as a clock. Recent work reported in the literature attests to the
  deep and manifold implications of the timeless picture of quantum mechanics.
  Research into this subject
  has led to the re-consideration of well-known foundational issues, such as Pauli's
  famous argument for the impossibility of a time observable in quantum mechanics \cite{LM2017}.
  New facets of time in quantum mechanics have been discovered, such as its
  basic connection with quantum coherence \cite{MS2019}. Interesting forays into relativistic scenarios
  have also been made, with the implementation of the PW scheme for Dirac \cite{DR2019}
  and scalar \cite{DMR2019} particles. A formalism akin to the one behind the timeless
  picture has led to the development of new, practical computational techniques
  for problems in quantum dynamics, which are reformulated as ground-state
  eigenvalue problems \cite{MPA2013}. Going beyond theoretical considerations,
  concrete experiments illustrating the timeless picture have been
  successfully conducted in recent years \cite{MBGGMG2014,MSVW2015,PRBGBRL2019}.

  As already mentioned, the system-clock entanglement is central to the timeless
  approach to quantum dynamics. However, the quantitative relation between
  quantum correlations and specific, dynamic-related aspects of the evolving
  system $R$ has received relatively little attention, with
  most efforts focusing on scenarios where the system-clock composite is in a pure state \cite{BRGC2016,BR2018,MVMP2019}. Our aim in this work is to explore the timeless
  picture of quantum dynamics for mixed global states of the bipartite system $R+C$.

  Motivations to study mixed states within this context are
 manifold. First, the system $R$ is, in general, itself composite. In realistic circumstances one may have
 access only to a subsystem $R_a$ of $R$ that, while weakly coupled to other parts of $R$, may nevertheless
 be entangled to them and, consequently, be in a mixed state. In this scenario the system $R=R_a+R_b$ has a total Hamiltonian ${\hat H}_R \approx {\hat H}_{R_a} + {\hat H}_{R_b}$, where ${\hat H}_{R_a}$ and ${\hat H}_{R_b}$  are the Hamiltonians of $R_a$ and $R_b$, respectively. If the state of $R$ is $\sigma_R$,
 the subsystem $R_a$ is described by the reduced, marginal state $\sigma_{R_a} = {\rm Tr}_{R_b} [\sigma_R ]$
 obeying the von Neumann equation
 \be \label{Greatvonneumann}
 \frac{d \sigma_{R_a}}{dt} = \frac{1}{i \hbar} [{\hat H_{R_a}},\sigma_{R_a}].
 \ee
 \noindent
 The subsystem $R_a$ is then, for all intents and purposes, our ``rest of the Universe".
 The PW approach, in its standard formulation (pure-state version), studies how the dynamics
 of an isolated system $R$ described by a pure state and obeying Schr\"odinger
 equation can be embedded into a stationary pure state of the $R+C$ system.
 It is legitimate and pertinent to inquire if the dynamics of a system $R_a$
 evolving according to the von Neumann's equation (\ref{Greatvonneumann}), which is the most general
  equation of motion for a closed quantum system, can similarly be embedded into
  a stationary \emph{mixed} state of $R_a + C$. One can, of course, circumvent the need to consider mixed states by implementing
 the pure-state version of the PW picture for the complete composite $R_a+ R_b + C$,
assumed to be in a pure state. But that procedure
 entails carrying the excess baggage of describing all the degrees of freedom
 of subsystem $R_b$, which may be inaccessible and irrelevant. Avoidance of that extra load
 is the main motivation for considering the PW approach for mixed states, which, by the way,
 coincides with the very reason for using the von Neumann equation
 (\ref{Greatvonneumann}) to study the dynamics of entangled, but dynamically-isolated,
  subsystems. In a cosmological context, the aforementioned picture is consistent with the one advanced by Bunyi and Hsu in \cite{BH2012}. According to these authors, the standard Big Bang cosmological
  model implies that a given subsystem  $R_a$  of the Universe is likely to be entangled to other subsystems with which $R_a$ is not currently interacting.

  Second, relevant motivations for developing a PW approach for mixed states are not limited to scenarios, as those discussed above, where mixed states describe subsystems of a composite quantum Universe that, as a whole, may be in a pure state. Indeed, the Universe itself (that is, the whole system $R+C$) may conceivable be in a mixed state \cite{P1986,GK1989}. This possibility was entertained by Page in \cite{P1986}, where a quantum description
  of the Universe was proposed which, in contrast with the celebrated Hartle-Hawking wave function, corresponds to a density matrix describing an impure quantum state. Physical effects depending on the degree of mixture of the density matrix of the Universe were considered by Gurzadyan and Kocharyan in \cite{GK1989}. In the present work, according to the above discussion, the expression ``rest of the Universe" may refer either to the ``total rest" $R$, or to an appropriate subsystem $R_a$. In the latter case we shall drop the subindex $a$.

  Additionally, considering the more general framework given by mixed states may also help in
  adapting the timeless PW picture to extensions or modifications of quantum mechanics motivated
  by research into the interface between quantum and gravitational phenomena. In this regard, we can mention   Deutch's proposal for a formulation of quantum mechanics in the presence of closed time-like curves (CTC),
  which explicitly requires density matrices describing mixed states, and cannot be formulated in terms of wave functions \cite{D1991}. Last, the analysis of mixed states in connection with the timeless approach  to quantum mechanics may shed new light on the problem of the ontological status of mixed states \cite{AA1999}.

  All the above motivations can be encompassed by a single aim: to formulate the PW picture in a fashion that incorporates the most general description of the dynamics of a closed quantum system, which is the one given by von Neumann's equation for the evolution of time-dependent mixed states. Therefore, in the present work   we advance a PW-like static scenario involving mixed global states of composite $R+C$. Our proposal
  is compatible with general time-dependent mixed states of the evolving system $R$, but otherwise keeps
  the main assumptions of the PW scenario. This mixed-state version of the PW approach constitutes a proof of principle showing that a consistent mixed-state PW scenario can be developed, and provides a testing ground to explore possible physical features of such a scenario,
   particularly in connection with quantum entanglement. We shall
   analyze quantitatively the entanglement between $R$ and $C$ and its relation with the emerging time evolution experienced by $R$. We investigate a quantitative indicator of entanglement, based on comparing an entropic measure evaluated on the global system with the corresponding entropic measure evaluated on the reduced state associated with $R$.
Using this indicator of entanglement we prove that (under the PW constraint of a definite total energy
of $R+C$ equal to zero) the composite system $R+C$ necessarily
has to be entangled for the system $R$ to exhibit dynamical evolution, meaning that other forms of non-classical correlations alone are not sufficient for time and evolution to arise. We establish an upper bound, as well as the asymptotic value, of the entanglement indicator, expressing these quantities in terms of an entropic measure of the spread of the energy probability distribution associated with the system $R$. We also investigate the connection between the entanglement present in the $R+C$ composite, and a measure of the energy uncertainty of the system $R$.

The paper is organized as follows. A brief summary of the timeless approach is given in
Section 2. The connection between time evolution and
entanglement for mixed states of the whole system $R+C$ is analyzed
in Section 3, on the basis of an appropriate entropic entanglement indicator.
An upper bound and the asymptotic limit of this quantity is discussed
in Section 4. Its connection with the energy dispersion of the system
is investigated in Section 5. Finally, some concluding remarks
are given in Section 6.

\section{Timeless Approach to dynamics for pure states of the system-clock composite}\label{EEvol}

As a starting point, we consider  a bipartite quantum system (the \emph{Universe} $U$)
comprising a \textit{clock} ($C$) and the \textit{rest} of the Universe ($R$).
The  Hilbert spaces corresponding to these two subsystems and to the total system
are, respectively, $\mathcal{H}_C$, $\mathcal{H}_R$, and $\mathcal{H}_U=\mathcal{H}_R\otimes
\mathcal{H}_C$. Global states of $R+C$ are spanned in a product orthonormal basis $\{\ket{\vect x}\otimes\ket{t}=\ket{\vect x}\ket{t}\}$ of $\mathcal{H}_U$, where $\{\ket{t}\}$ and $\{\ket{\vect x}\}$ are orthonormal basis of $\mathcal{H}_C$ and $\mathcal{H}_R$, respectively.
The continuous label $t \in \mathbb{R}$ characterizing the basis states of $\mathcal{H}_C$
corresponds to the eigenvalues of an observable $\mathcal{\hat T}$ associated with
the position of the clock's hands, so that $\mathcal{\hat T}\ket{t}=t\ket{t}$.
Likewise, the label $\vect x$ characterizing the basis states of $\mathcal{H}_R$
represents the position, or any  other degrees of freedom, of the particle or particles
 constituting the system $R$. Throughout the paper we will assume that
$\vect x$ is a continuous variable, yet it may also denote a discrete one,
provided integrals are properly substituted by discrete sums.

 To analyze the behavior of the complete system during a finite time interval $[0,T]$,
 we assume that $U$ is in the pure state
\be \label{Psi}
 |\Psi \rangle \, = \frac{1}{\sqrt{T}}\, \int \,  \Psi(\vect x,t) \, |\vect x\rangle  \,  |t \rangle  \, d\vect x  \, dt,
\ee
described by a wave function $\Psi(\vect x,t) = (\langle \vect x| \langle t|) |\Psi \rangle$ that is spatially
normalized, $\int  |\Psi(\vect x,t)|^2 d\vect x  = 1$.
The state (\ref{Psi}) is then properly normalized, both spatially and temporally,
\be \label{PsiNorm}
 \langle \Psi|\Psi \rangle \, = \frac{1}{T}\, \int^T_0 \, \underbrace{\left(\int |\Psi(\vect x,t)|^2  \, d\vect x \right)}_{=1} \, dt=1.
\ee

The state of $R$ for a given configuration of the clock's hands
(that is, for a particular value of $t$) is
described by the \textit{Everett relative} state \cite{E1957}
\be \label{relastate}
|\Phi_t  \rangle = \langle t|\Psi\rangle=\frac{1}{\sqrt{T}}\int \,  \Psi(\vect x,t) \, |\vect x\rangle  \,   d\vect x=
\frac{1}{\sqrt{T}}|\,\widetilde{\Phi}_t  \rangle,
\ee
obtained by projecting $|\Psi \rangle$ onto $\ket{t}$. In (\ref{relastate}),
$|\widetilde{\Phi}_t  \rangle$ stands for the \emph{normalized} relative state, satisfying
\be \label{norm2}
\langle \widetilde{\Phi}_t|\widetilde{\Phi}_t \rangle\, = T\,\langle \Phi_t|\Phi_t \rangle=1.
\ee
Restricting our analysis to the finite time interval $[0,T]$ corresponds to consider a part of the history of the Universe that, from the standard time-based viewpoint, is perceived as having a duration $T$.
Quantum states normalized within the range $[0,T]$ result from projecting the state of the Universe onto the subspace spanned by the eigenstates of $\mathcal {\hat T}$ having eigenvalues $t\in [0,T]$. These states can be regarded as the result of post-selecting the measurement value $1$ when measuring the observable (projector) $\Pi = \int_0^T \, dt \, |t\rangle \langle t|$.

Within the timeless formalism it is assumed that
\be \label{similwhellerdewit}
\hat{H}_U\ket{\Psi}=0,
 \ee

 \noindent
 where $\hat{H}_U$ is the total Hamiltonian $\hat{H}_U=\hat H_R\otimes \mathbb{I}_C+\mathbb{I}_R \otimes \hat H_C$, with $\hat H_R$ an arbitrary Hamiltonian of $R$, and $\hat H_C$ the Hamiltonian of the clock.
Notice that a good clock, in order to appropriately keep track of time, has to be dynamically
 isolated, and should not be perturbed by interactions with other systems. The absence of interaction
 between $R$ and $C$ also plays a crucial role in guaranteeing the uniqueness of the $R|C$
 bipartition of the complete system $U$ \cite{MV2017}, solving the ambiguity problem raised in \cite{AI2008}.
 As discussed in \cite{MV2017}, we consider here an \emph{ideal} clock, that does not interact at all with $R$.  Furthermore, it is considered that the clock's observable $\mathcal{\hat T}$ and the Hamiltonian $\hat H_C$ satisfy the commutation relation $[\mathcal{\hat T},\hat H_C]=i\hbar$. Under these conditions, it
 follows from (\ref{similwhellerdewit}) \cite{V2014} that
the relative state $|\Phi_t  \rangle$ (whether normalized or not) obeys the Schr\"odinger equation,
\be \label{Schro}
i\hbar \frac{\partial}{\partial t} |\Phi_t  \rangle\, = \, \hat{H}_R |\Phi_t \rangle.
\ee
We thus see that the usual dynamical scenario ---embodied in the (time-dependent) Schr\"{o}dinger equation--- ensues from the static image of the non-evolving state $\ket{\Psi}$. The important point here to be noticed, is that
the evolution emerges if and only if $C$ and $R$ are entangled. Otherwise $\Psi(\vect x,t)$ factorizes as $\Psi(\vect x,t) = \Psi_C(t) \Psi_R(\vect x)$, $\Psi_R(\vect x)$ is an eigenstate of
$\hat H_R$, and therefore $\Psi(\vect x,t)$ is a stationary (non-evolving) state. Such intimate relation between entanglement and time evolution has been explored previously \cite{BRGC2016,BR2018,MVMP2019}
in this pure-case scenario. In what follows we will analyze the more general case of mixed states, and show that the relation still holds.

Before ending this section, let us add a few comments regarding the eigenvalue spectra of the
observables $\mathcal {\hat T}$ and $\hat H_C$. Both of them have continuous spectra. The allowed
eigenvalues of $\hat H_C$, however, are restricted to a discrete set, because the complete system
is assumed to be in a zero-energy eigenstate of the total Hamiltonian $\hat H_R\otimes \mathbb{I}_C+\mathbb{I}_R \otimes \hat H_C$.
Indeed, if $\hat H_R$ has a discrete spectrum consisting of the eigenvalues $\{E_n\}, \,\,\, n=0, 1, 2, \ldots$ (as we assume here), then the set of allowed eigenvalues of $\hat H_C$ is discrete too:
these eigenvalues, and their associated eigenstates, are  $\{- E_n\}$ and $\{ e^{-iE_n t /\hbar}\}$,
respectively. We have, therefore, an \emph{effectively} discrete Hilbert space for $C$, spanned by this discrete set of states.
This resembles what happens with the operators $\hat x$ and $\hat p$ of a particle in a finite box. The
commutation relation satisfied by these operators coincides with the one satisfied by $\mathcal {\hat T}$ and $\hat H_C$. Imposing appropriate boundary conditions on the walls of the box yields a discrete set $\{p_n\}$ of accessible eigenvalues for $\hat p$, with a corresponding discrete set of eigenvalues $\{ e^{-ip_n x /\hbar}\}$, that span the Hilbert space of the system.
Of course, the above similarity is only formal, since in the present situation the (effective) discrete spectra of $\hat H_C$ results from the constraint of total zero energy of the $C+R$ composite,
instead of arising from boundary conditions.
\\
\section{Evolution and entanglement for mixed states of the system-clock composite}\label{sec:mix}

In order to extend the above ideas beyond scenarios corresponding to pure
global states of the $R+C$ system, we shall assume a mixed global
state $\rho$ that is stationary under the dynamics determined by the
total Hamiltonian $\hat{H}_U$, and  has a definite total energy equal to $0$.
That is, we shall assume that $\langle \hat{H}_U \rangle = {\rm Tr}\,(\rho \, \hat{H}_U)= 0$
and $\langle \hat{H}_U^2 \rangle - \langle \hat{H}_U \rangle^2 =
{\rm Tr}[ \rho \,( \hat{H}_U - \langle \hat{H}_U \rangle)^2 ]= 0$.
The state $\rho$ is then of the form
\be \label{mixeduniverse}
\rho = \sum_j p_j |\Psi_j\rangle \langle \Psi_j |,
\ee
\noindent
where $p_j\geq 0$ for all $j$, $\sum_j p_j =1$, and $\{|\Psi_j\rangle\}$ is a set of
stationary pure states of $U$ with $\hat{H}_U |\Psi_j\rangle = 0$ for all $j$.
The density matrix (\ref{mixeduniverse}) describes thus a statistical mixture of the pure states
$ |\Psi_j\rangle$ with (probability) weights $p_j$.

Using the same notation as in the previous section, we have
\be \label{Psii}
|\Psi_j\rangle = \frac{1}{\sqrt{T}}\int \Psi_j(\vect x,t) |\vect x\rangle |t\rangle \,d\vect x \,dt,
\ee
\noindent
and the corresponding relative state
\be\label{Phit}
|\Phi_{j,t}\rangle =\langle t|\Psi_j\rangle=\frac{1}{\sqrt{T}}\int \Psi_j(\vect x,t) |\vect x\rangle \,d\vect x=\frac{1}{\sqrt{T}}|\widetilde\Phi_{j,t}\rangle,
\ee
\noindent
where $
|\widetilde\Phi_{j,t}\rangle = \sqrt{T}|\Phi_{j,t}\rangle
$
stands for the normalized relative states ($\langle\widetilde\Phi_{j,t} |\widetilde\Phi_{j,t}\rangle=1$).

Now, in this case the Everett relative state, describing the state of $R$ given that the clock's hands state is $\ket t$, is obtained according to
\begin{eqnarray} \label{RofUmix}
 \sigma_{R,t}=\frac{\textrm{Tr}_{C}\,(\ket{t}\bra{t}\rho)}{\textrm{Tr}\,(\ket{t}\bra{t}\rho)}&=&T\sum_j p_j \ket{\Phi_{j,t}}\bra{\Phi_{j,t}}\nonumber\\
 &=&\sum_j p_j |\widetilde\Phi_{j,t}\rangle \langle \widetilde\Phi_{j,t} |.
\end{eqnarray}
This is a mixture of the states $\ket{\Phi_{j,t}}$, each of which satisfies the Schr\"odinger equation (\ref{Schro}). Therefore, the relative state of $R$ satisfies the von Neumann equation,
\be \label{vonN}
\frac{d}{dt} \sigma_{R,t} = \frac{1}{i\hbar} [{\hat H_R},\sigma_{R,t}].
\ee
We thus verify that the quantum dynamical equations of $R$ are recovered also in the mixed state case.

In order to investigate the relation between the evolution and the entanglement in this more general scenario, we shall use an entanglement criteria based on the reduced, marginal, density matrix $\rho_R$ of the system $R$, obtained by taking
the partial trace over $C$ of the global density matrix $\rho$:
\ben \label{rho_Csigma}
\rho_R = \textrm{Tr}_C\, \rho&=&\int_0^{T}   \langle t |\rho |  t \rangle\,dt\cr
&=&\frac{1}{T}\int_0^{T} \sigma_{R,t}\, dt=\overline{\sigma_{R,t}},
\een
where $\overline{(\cdot)}$ denotes the time average $\overline{(\cdot)}=\frac{1}{T}\int_0^{T} (\cdot)\, dt$.
It is worth to emphasize that the density
matrices  $\sigma_{R,t}$ and $\rho_R$, though both referring to system $R$, represent different states.
The former represents the state of $R$ \textit{conditioned} to the state $\ket{t}$ of the clock, and is a mixed state
that evolves unitarily as a function of the parameter $t$. On the other hand, the (in general) mixed state $\rho_R$
is obtained through taking, on the global state of $R+C$, the partial trace over the degrees
of freedom of $C$. It represents a time-averaged state (over the interval $[0,T]$) and does not  depend on $t$.

Now, the entropies $S[\rho]$ and $S[\rho_R]$ of the global ($\rho$) and the marginal ($\rho_R$) density matrices, respectively, provide an entanglement criterion for the global
state as follows (see \cite{NK2001,BPMP2002,HH1996,TLB2001,RC2002})
\be \label{crit}
S[\rho_R]>S[\rho]\Rightarrow \rho \textrm{  is entangled}.
\ee
That is, if we have less information about the subsystem $R$ than information about the composite system $R+C$, then $R$ and $C$ are entangled. This entropic entanglement criterion can be implemented irrespective of the particular entropic measure used. Possible choices are von Neumann entropy, or
the \textit{linear entropy} defined, for a generic density matrix $\varrho$, as
\be \label{SLgen}
S_L[\varrho]\equiv1-\textrm{Tr}\,\varrho^2.
\ee
Since this latter has some computational advantages, we will choose it for our calculations, and thus compare $S_L[\rho]$ with $S_L[\rho_R]$. Our entanglement indicator is thus
\be
\Delta S\equiv S_L[\rho_R]-S_L[\rho],
\ee
in terms of which the entanglement criterion reads
\be \label{critDelta}
\Delta S>0\Rightarrow \rho \textrm{  is entangled}.
\ee

The linear entropy of the global state $\rho$ is given by
\ben \label{SLrho}
S_L[\rho] &=& 1 - \textrm{Tr}\,\rho^2
 = 1 - \sum_{jk} p_j p_k |\langle \Psi_j | \Psi_k \rangle |^2 \cr
&=& 1 - \sum_{jk} p_j p_k \Bigl| \frac{1}{T}\int_0^T
 \langle \widetilde\Phi_{j,t} |  \widetilde\Phi_{k,t} \rangle \,dt\,\Bigr|^2.
\een
Since the inner product $ \langle \widetilde \Phi_{j,t} | \widetilde \Phi_{k,t} \rangle $ is invariant under the unitary evolution determined by the Schr\"odinger equation, we can substitute $\langle \widetilde\Phi_{j,t} |  \widetilde\Phi_{k,t} \rangle = \langle \widetilde\Phi_{j,0} |  \widetilde\Phi_{k,0} \rangle$ in the above equation and get
\ben \label{prometuno}
S_L[\rho] &=&  1 - \sum_{jk} p_j p_k | \langle \widetilde\Phi_{j,0} | \widetilde \Phi_{k,0} \rangle  |^2 \cr
&=& 1 - \sum_{jk} p_j p_k | \langle \widetilde\Phi_{j,t} | \widetilde \Phi_{k,t} \rangle  |^2 \cr
&=&1 - \textrm{Tr}\, \sigma_{R,t}^2 =S_L[\sigma_{R,t}]\cr
&=&\overline{S_L[\sigma_{R,t}]},
\een
\noindent
where the last equality is due to the fact that (as follows from the first two lines) $S_L[\sigma_{R,t}]$ is a time-independent quantity.

As for the linear entropy of the marginal state $\rho_R$, Eq. (\ref{rho_Csigma}) gives
\be \label{prometdos}
S_L[\rho_R] =
S_L \left[ \overline{\sigma_{R,t}}\right].
\ee
It follows from the above expressions that
comparing the entropies of $\rho$ and $\rho_R$ amounts to compare the (time) average
entropy of $\sigma_{R,t}$ with the entropy of the (time) average of $\sigma_{R,t}$.

Now, given a time-dependent density matrix $\varrho_t$ and
a concave function $f(x)$, the following inequality holds:
\be \label{ineq}
\textrm{Tr} \left[f(\overline{\varrho_t})\right] \ge \overline{\textrm{Tr}\,[f(\varrho_t)]},
\ee
with the equality satisfied only if  $\varrho_t$ is constant in time \cite{W1978}.
In particular, for $f(x)=x-x^2$, we get $S_L(\varrho_t)=\textrm{Tr}\,f(\varrho_t)$, the inequality (\ref{ineq}) leads to
\be \label{ineqSLok}
S_L[\overline{\varrho_t}] \geq \overline{S_L[\varrho_t]},
\ee
and therefore, putting $\varrho_t=\sigma_{R,t}$ it follows from (\ref{prometuno})-(\ref{prometdos}) that
\be \label{entrinequa}
S_L[\rho_R] \ge S_L[{\rho}],
\ee
\noindent
with the two entropies appearing in the above equation being equal only if there
is no time evolution. In other words, $\Delta S$ vanishes only if $\sigma_{R,t}$
does \emph{not} evolve in time, that is, $\Delta S=0\Rightarrow$ no time evolution.

Since Eq. (\ref{entrinequa}) holds for \emph{all} states belonging to the subspace spanned by the eigenstates of zero energy of the total Hamiltonian $\hat H_U$, it follows from the criterion (\ref{critDelta}) that \emph{all} these states are entangled, provided $\sigma_{R,t}$ evolves in time. In other words, if the (relative) state of $R$ changes with the ticking of the clock's hands, then $R$ is necessarily entangled with $C$. Put another way, in the absence of entanglement between the clock and the system $R$, the state of $R$ remains independent of $t$, and no evolution occurs. This means that, under the conditions of our
 mixed-state PW scenario, quantum correlations other than entanglement that may be present in mixed states, such as quantum discord, are not enough for dynamics and the flow of time to arise. Therefore, the study of mixed states within the timeless approach to quantum dynamics provides further evidence for the intimate link existing between entanglement and evolution.\\

Finally, it is interesting to ask whether for the type of density matrices arising in the
present PW context, the condition $\Delta S=0$ implies that the entanglement between $R$ and $C$
vanishes (this would amount to state that an entangled state implies $\Delta S>0$, and consequently the criterion (\ref{critDelta}) would be not only sufficient but also necessary). For the special case of pure states of the $R+C$ system the answer is yes. For mixed states the situation is more subtle.
In such case, as we have just seen, $\Delta S=0$ implies that the
relative state of $R$ conditional to a given value of $t$, does not depend
on this $t$-value (that is, $R$ does not evolve). However, this condition does not seem to imply that the joint density matrix
of $R+C$ is non-entangled. It might
happen that there are entangled joint states of $R+C$ for which
$R$ does not evolve. The existence or not of such states remains
an open question, certainly worth of further investigation.

\section{Upper Bound and Asymptotic Limit of the entanglement indicator}\label{sec:upperasym}

Now we shall determine an upper bound for the indicator $\Delta S$
of entanglement between the system and the clock, and also its asymptotic
limit for large lengths of the interval $[0,T]$ within
which the joint state of the system-clock composite is defined.
We consider a $d$-level system with a Hamiltonian $\hat H_R$ having
 eigenstates $\{\ket{0},\ket{1},\dots,\ket{d-1}\}$ with corresponding
 eigenvalues $\{E_0,E_1,\dots,E_{d-1}\}$. The relative state $\sigma_{R,t}$  evolves according to Eq. (\ref{vonN}), and its matrix elements in the basis $\{\ket{n}\}$ (with $n=0,\dots,d-1$) can thus be written as
\be\label{sigmaqdit}
\sigma_{nm}(t)\equiv \langle n|\sigma_{R,t}|m\rangle=e^{-i(E_n-E_m)t/\hbar}\,\sigma_{nm}(0).
\ee
The matrix elements of the reduced state $\rho_R$ are given, according to Eqs. (\ref{rho_Csigma}) and (\ref{sigmaqdit}), by
\begin{gather}
\label{rhoRqdit}
\langle n|\rho_{R}|m\rangle= \langle n|\overline{\sigma_{R,t}}|m\rangle\nonumber\\
= \sigma_{nm}(0)e^{i(E_n-E_m) T/2\hbar}\,\textrm{sinc}\,[(E_n-E_m)T/2\hbar],\label{rhoRqdit}
\end{gather}
with sinc$\,x=x^{-1}\sin x$.

From these expressions the linear entropies $S_L[\rho]$ and $S_L[\rho_R]$ can be computed directly as follows
\begin{eqnarray}\label{Srho}
S_L[\rho]=S_L[\sigma_{R,t}]&=&1-\textrm{Tr}\,\sigma_{R,t}^2=1-\sum_{nm}\sigma_{nm}\sigma_{mn}\nonumber\\
&=&1-\sum_{nm}|\sigma_{nm}(0)|^2,
\end{eqnarray}
and
\begin{eqnarray}\label{SrhoR}
S_L[\rho_R]&=&1-\textrm{Tr}\rho^2_R=1-\sum_{nm} \langle n|\rho_{R}|m\rangle \langle m|\rho_{R}|n\rangle\nonumber\\
&=&1-\sum_{nm}|\sigma_{nm}(0)|^2 \textrm{sinc}^2\,(\omega_{nm}T/2),
\end{eqnarray}
where $\omega_{nm}=|E_n-E_m|/\hbar$. Decomposing the sum in (\ref{SrhoR}) into those terms for which $\omega_{nm}=0$ and those for which $\omega_{nm}\neq 0$ we get
\begin{eqnarray}\label{SrhoRbis}
S_L[\rho_R]&=&\Big(1-\sum_{\substack{nm \\ (\omega_{nm}=0)}}|\sigma_{nm}(0)|^2 \Big)-\nonumber\\
&&-\sum_{\substack{nm \\ (\omega_{nm}\neq0)}}|\sigma_{nm}(0)|^2\textrm{sinc}^2\,(\omega_{nm}T/2).
\end{eqnarray}
The entanglement indicator is thus
\begin{eqnarray}
\Delta S&=& S_L[\rho_R]-S_L[\rho]\nonumber\\
&=&\sum_{nm}|\sigma_{nm}(0)|^2[1-\textrm{sinc}^2\,\,(\omega_{nm}T/2)]\nonumber\\
&=&\sum_{\substack{nm \\ (\omega_{nm}\neq0)}}|\sigma_{nm}(0)|^2[1-\textrm{sinc}^2\,\,(\omega_{nm}T/2)]\label{difcom}\\
&\leq&\sum_{\substack{nm \\ (\omega_{nm}\neq0)}}|\sigma_{nm}(0)|^2,\label{dif}
\end{eqnarray}
and its maximum value ---which coincides with its asymptotic value when $T\rightarrow \infty$--- is
\be \label{max}
\Delta S_{\max}=\sum_{\substack{nm \\ (\omega_{nm}\neq0)}}|\sigma_{nm}(0)|^2.
\ee
Notice that the condition $\omega_{nm}\neq 0$ introduced above is not necessarily equivalent to $n\neq m$, due to possible degeneracies of the energy eigenvalues. The label $n$ should therefore be understood as
representing a (possibly compound) index containing all the quantum numbers required to completely characterize the eigenstates of $\hat H_R$. The set of possible values of this index (even if it is compound)
is at most denumerable, and can thus be regarded as ordered in the sequence $n = 0, 1, \ldots$.

Now, let us denote with $\sigma_{R|M}$ the state of $R$ obtained when a \emph{non-selective}
energy measurement is performed on $R$, that is
\be \label{sigmaE}
\sigma_{R|M}=\sum_Ep_E\sigma_{R|E}=\sum_E\Pi_E \sigma_{R,t} \Pi_E,
\ee
where $p_E={\rm Tr}\, (\Pi_E \sigma_{R,t})$ is the
probability of obtaining the result $E$ when measuring the energy of $R$
when it is in the state $\sigma_{R,t}$, $\sigma_{R|E}=\Pi_E \sigma_{R,t} \Pi_E/p_E$ is the (collapsed) state of $R$ obtained when the
energy measurement yields the result $E$,
and $\Pi_E = \sum_{\substack{n\\ (E_n=E)}} |n\rangle\langle n |$
is the projector onto the subspace spanned by the degenerated eigenstates $|n\rangle$ that correspond to
the same energy eigenvalue $E$. The projector satisfies
\ben
\Pi_E\Pi_{E'}&=&\sum_{\substack{nm\\ (E_n=E),(E_m=E')}} |n\rangle\langle n |m\rangle\langle m |\cr
&=&\delta_{EE'}\sum_{\substack{n\\ (E_n=E)}} |n\rangle\langle n |=\delta_{EE'}\,\Pi_E,\label{PiEdelta}
\een
so that
\be \label{sigmaEdelta}
\sigma_{R|E}\sigma_{R|E'}=\sigma^2_{R|E}\delta_{EE'}.
\ee
 Taking into account the second equality in (\ref{sigmaE}) and (\ref{PiEdelta}), we get for the linear entropy of the state $\sigma_{R|M}$,
\ben
S_L[\sigma_{R|M}]&=&1-\textrm{Tr}\,\sigma_{R|M}^2\cr
&=&1-\sum_{E}\textrm{Tr}\,(\sigma_{R,t}\Pi_E\sigma_{R,t}\Pi_E)\cr
&=&1-\sum_E\sum_{\substack{nm \\ (E_n=E_m=E)}}|\sigma_{nm}(0)|^2\cr
&=&1-\sum_{\substack{nm \\ (\omega_{nm}=0)}}|\sigma_{nm}(0)|^2,
\een
which combined with Eqs. (\ref{Srho}) and (\ref{max}) leads to
\be \label{deltamax}
\Delta S_{\max}=S_L[\sigma_{R|M}]-S_L[\sigma_{R,t}].
\ee
This relation shows that the asymptotic value, when $T\rightarrow \infty$, of the entanglement indicator is given by the difference between the entropy of the state of $R$ after and before a non-selective energy measurement is performed.

The entropy $S_L[\sigma_{R|M}]$ bears information regarding the possible states $\sigma_{R|E}$ that can be obtained after an energy measurement, and also regarding the energy probability distribution $\{p_E\}$. Such information can be extracted by recourse to the first equality in Eq. (\ref{sigmaE}) and to (\ref{sigmaEdelta}), obtaining
\ben
S_L[\sigma_{R|M}]&=&1-\textrm{Tr}\,\sigma_{R|M}^2\cr
&=&1-\textrm{Tr}\,\sum_{EE'}p_E\,p_{E'}\,\sigma_{R|E}\,\sigma_{R|E'}\cr
&=&1-\sum_Ep^2_E\,\textrm{Tr}\,\sigma^2_{R|E}\cr
&=&S_L[\{ p_E \}] + \sum_E p_E^2 \, S_L[\sigma_{R|E}],
\een
where $S_L[\sigma_{R|E}]$ stands for the linear entropy associated to the state $\sigma_{R|E}$, and $S_L[\{ p_E \} ] = 1 - \sum_E p_E^2$ is the linear entropy corresponding to the energy probability distribution $\{ p_E \}$.

It is instructive to consider particular cases of the bound $(\ref{deltamax})$. When the spectrum of $\hat H_R$ has no degeneracy, one has $S_L[\sigma_{R|E}]=0$, hence $S_L[\sigma_{R|M}]$ becomes $S_L[\{ p_E \} ] $, and the upper bound
reduces to
\be
(\Delta S_{\max})|_\textrm{non-degenerate} = S_L[\{ p_E \} ] - S_L[\sigma_{R,t}].
\ee
It is also particularly interesting to see what happens if the global state $\rho$ is pure, so that $S_L[\rho]=S_L[\sigma_{R,t}]=0$. In this case also $\sigma_{R|E}$ is a pure state, whence $S_L[\sigma_{R|E}]=0$, and again $S_L[\sigma_{R|M}]=S_L[\{ p_E \} ] $.
Consequently, for pure states one recovers the expression \cite{MVMP2019}
\be \label{deltamaxP}
(\Delta S_{\max})|_\textrm{pure}=S_L[\{ p_E \} ],
\ee
meaning that the upper bound of the $S_L$-based indicator of entanglement
is given by the spread of the energy probability distribution $p_E$, as measured by its linear entropy.
This is no longer the case for mixed states. In an extreme case, for example, in which $\rho$ is diagonal in the energy eigenbasis one has $\sigma_{nm}\sim \delta_{nm}$, and Eq. (\ref{max}) leads straightforward to
\be \label{deltamaxD}
(\Delta S_{\max})|_\textrm{diagonal}=0,
\ee
meaning that diagonal states do not evolve in time (see below Eq. (\ref{entrinequa})).
Now, when $\rho$ is diagonal in the energy eigenbasis, all
the spread in the energy probability distribution is purely classical, whereas for pure states (that are no energy eigenstates) all the spread in the energy probability distribution is of a quantum
nature. These observations, together with Eqs. (\ref{deltamaxP}) and (\ref{deltamaxD}), indicate that only the quantum component of the spread in the energy probability distribution contributes to the upper bound
of the system-clock entanglement.

\subsection{An example. The qubit case}

As an illustration of our previous results we consider now a qubit (two-level) system with a Hamiltonian
$\hat H_R$ having eigenstates $\ket{0}$ and $\ket{1}$, with corresponding eigenvalues $E_0$ and $E_1$.
Following Eq. (\ref{sigmaqdit}), the relative state $\sigma_{R,t}$ in the basis $\{\ket{0},\ket{1}\}$ reads
\begin{eqnarray}\label{sigmaqubit}
\sigma_{R,t}=    \begin{pmatrix}
\sigma_{00}(0)&e^{i\epsilon t/\hbar}\,\sigma_{01}(0)\\
e^{-i\epsilon t/\hbar}\,\sigma^{*}_{01}(0)&1-\sigma_{00}(0)\\
\end{pmatrix},
\end{eqnarray}
where we wrote $\epsilon=E_1-E_0$. The reduced density matrix $\rho_R$ is given, according to Eq. (\ref{rhoRqdit}), by
\begin{eqnarray}\label{rhoRqubit}
\rho_{R}= \overline{\sigma_{R,t}}=   \begin{pmatrix}
\sigma_{00}(0)&\sigma_{01}(0)e^{ix}\,\textrm{sinc} \,x\\
\sigma^{*}_{01}(0)e^{-ix}\,\textrm{sinc} \,x&1-\sigma_{00}(0)\\
\end{pmatrix},
\end{eqnarray}
with $x= \epsilon T/2\hbar$. The corresponding linear entropies are (see Eqs. (\ref{Srho}) and (\ref{SrhoR}))
\be
S_L[\rho_R]=2\Big\{\sigma_{00}(0)[1-\sigma_{00}(0)]-|\sigma_{01}(0)|^2 \textrm{sinc}^2\,x\Big\},
\ee
and
\begin{eqnarray}
S_L[\rho]&=&S_L[\sigma_{R,t}]\nonumber\\
&=&2\Big\{\sigma_{00}(0)[1-\sigma_{00}(0)]-|\sigma_{01}(0)|^2\Big\}.
\end{eqnarray}
The entanglement indicator is thus
\be \label{DeltaSqubit}
\Delta S=2|\sigma_{01}(0)|^2(1-\textrm{sinc}^2\,x),
\ee
which is greater than zero for $x>0$, provided $\sigma_{01}(0)\neq 0$. That is, for any nonzero $T$, the evolution of $R$ reflects its entanglement with the clock.

Basic features of the connection between the evolution of the qubit and
its entanglement with the clock can be appreciated in Figure 1.
The dependence of the entropies $S_L[\rho_R]$  and $S_L[\rho]$ on the
parameter $\sigma_{01}(0)$ is depicted in the left panel of the figure.
The entanglement indicator $\Delta S$ as a function of
$x= \epsilon T/2\hbar$ for two values of the parameter
$\sigma_{01}(0)$, is shown on the right panel. Notice that the entanglement indicator $\Delta S$ approaches its asymptotic limit rather quickly, becoming close to its upper bound already at values of $T$ corresponding to $\epsilon T/2\hbar \approx 2$.

\begin{figure}[h]
\begin{center}
\vspace{0.5cm}
\includegraphics[width=0.238\textwidth]{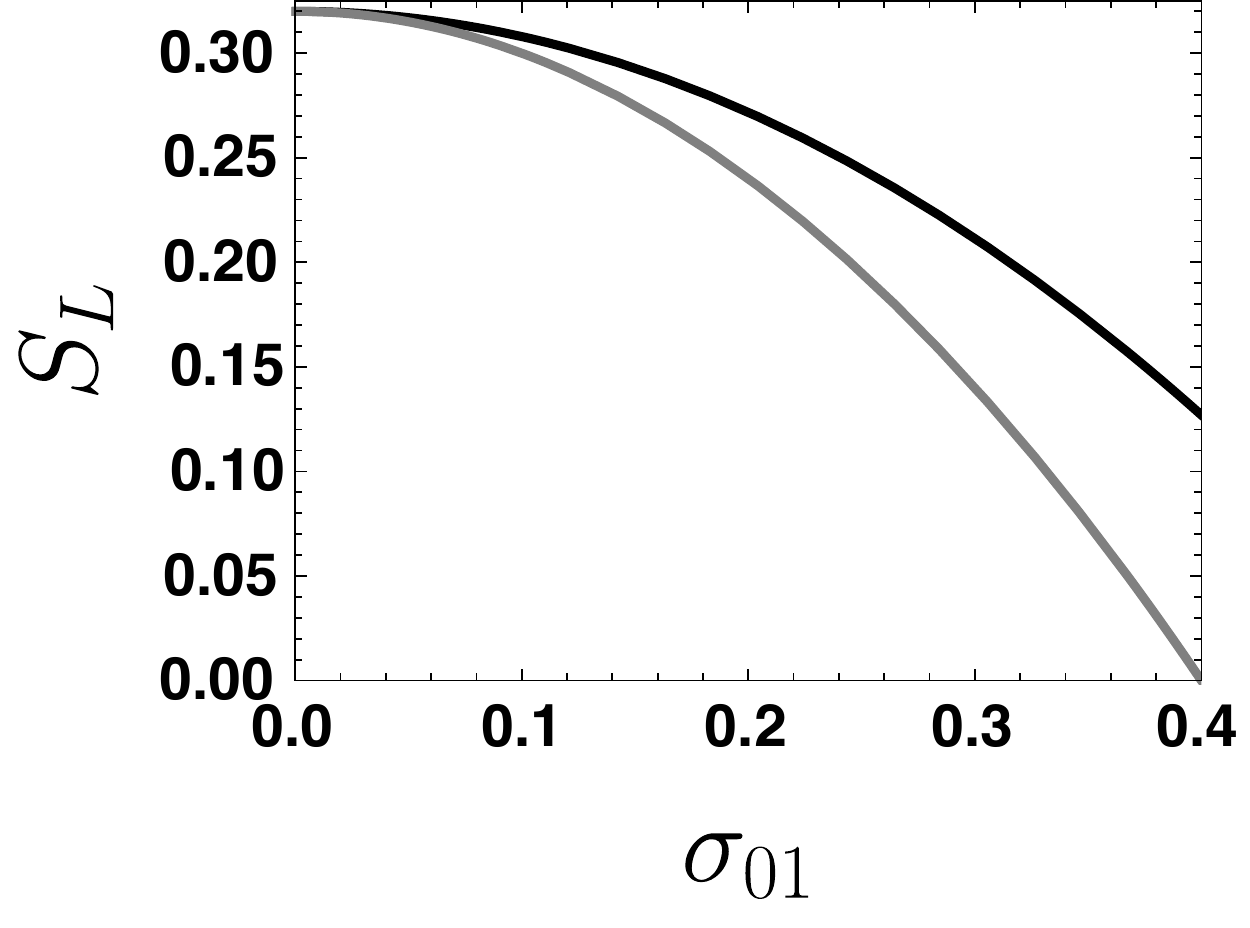}
\includegraphics[width=0.238\textwidth]{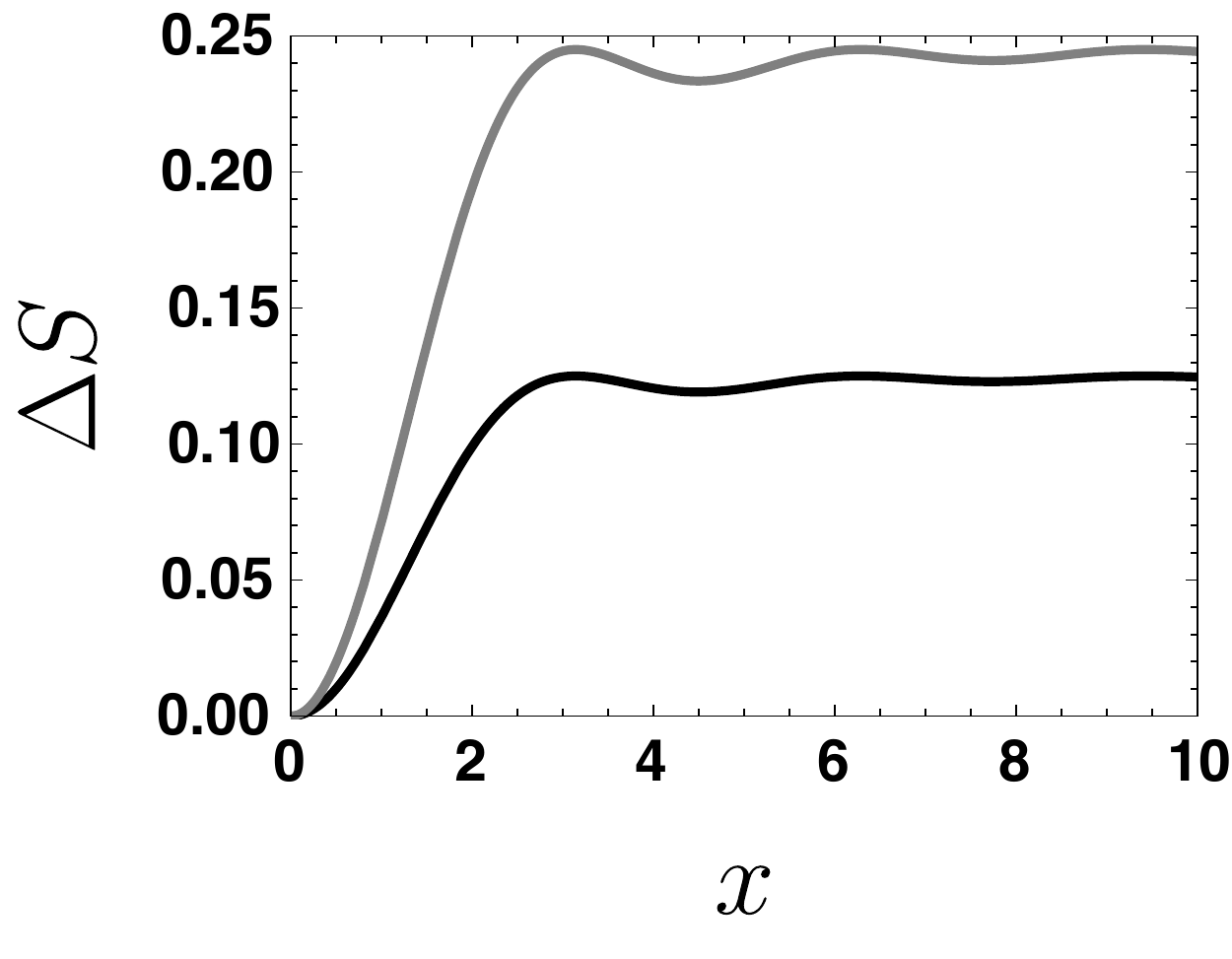}
\caption[]{\label{fig:thefig} Left panel: $S_L[\rho_R]$ (black) and $S_L[\rho]$ (grey) as a function of $\sigma_{01}(0)$ for a qubit state, setting $\sigma_{00}=0.2$ and $x=1.2$.  Right panel: $\Delta S$  as a function of $x$, setting $\sigma_{00}=0.2$ and $\sigma_{01}(0)=0.25$ (black), 0.35 (grey).
All depicted quantities are dimensionless.}
\end{center}
\end{figure}

\section{Relation Between the Entanglement Indicator and Energy Dispersion}\label{sec:energy}

   As we have seen in the previous sections, the entanglement between the system $R$ and the clock $C$
   is  linked to the time evolution of $R$. On the other hand, the evolution of a quantum system
   is closely related to the system's energy uncertainty. Consequently, there has to be a connection between
   the energy uncertainty of $R$, and the entanglement between $R$ and $C$. In this section we shall
   investigate such connection for mixed joint states of $R+C$.

By recourse  to Eq. (\ref{difcom}) and to the Taylor series of the $\textrm{sinc}$ function
 \begin{eqnarray}\label{sinc}
        \textrm{sinc} \,z=\sum^{\infty}_{l=0}\frac{(-1)^lz^{2l}}{(2l+1)!},
 \end{eqnarray}
it can be verified that to lowest order in $T$ the entanglement indicator
$\Delta S$ is

\ben \label{low}
\Delta S &=& \frac{T^2}{12 \hbar^2} \sum_{nm} \, |\sigma_{nm}(0)|^2 \left(  E_n - E_m \right)^2\cr
&=&- \frac{T^2}{12 \hbar^2} {\rm Tr}\left( [\hat H_R,\sigma_{R,t}]^2 \right) .
\een

\noindent
Here we face a situation similar to the one analyzed in the previous Section,
but now referred to the energy dispersion
\ben \label{Edisp}
\sigma^2_E&\equiv&  \langle \hat H_R^2 \rangle - \langle \hat H_R \rangle^2\cr
&=&{\rm Tr}\,(\hat H_R^2\sigma_{R,t})-{\rm Tr}^2\,(\hat H_R\sigma_{R,t})
\een
 instead of the spread in the energy probability distribution, as measured by $S_L[\{ p_E \} ]$. This can be seen as follows. For pure states $\sigma_{R,t}=|\tilde{\Phi}_t\rangle\langle\tilde{\Phi}_t|$ we have
\ben \label{trpure}
{\rm Tr}\left( [\hat H_R,\sigma_{R,t}]^2 \right)&=&{\rm Tr}\left( [\hat H_R,|\tilde{\Phi}_t\rangle\langle\tilde{\Phi}_t|]^2 \right)\cr
 &=&2(\langle\tilde{\Phi}_t|\hat H_R|\tilde{\Phi}_t\rangle)^2-2\langle\tilde{\Phi}_t|\hat H_R^2|\tilde{\Phi}_t\rangle\cr
 &=&-2\sigma^2_E,
\een
and we obtain the expression
\be \label{pure}
(\Delta S)|_\textrm{pure}=\frac{T^2}{6 \hbar^2}\sigma_E^2
\ee
relating, for pure states of $R+C$, the lowest-order expansion of the $S_L$-based  entanglement indicator (describing its behavior for short-time intervals), with the energy dispersion. In the other extreme situation, for mixed states that are diagonal in the basis of eigenvectors of $\hat H_R$, one has $[\hat H_R,\sigma_{R,t}]=0$, and Eq. (\ref{low}) gives
\be \label{diagonal}
(\Delta S)|_\textrm{diagonal}=0.
\ee
Equations (\ref{pure}) and (\ref{diagonal}) are analogous to Eqs. (\ref{deltamaxP}) and (\ref{deltamaxD}). As happens with the spread of the energy probability distribution, the energy dispersion
has both classical and quantum components. For pure states, all the energy dispersion is of quantum nature, whereas for mixed states that are diagonal in an energy eigenbasis, it
is purely classical. Thus, the quantity
\be
{\mathcal D}\equiv-{\rm Tr}\left([\hat H_R,\sigma_{R,t}]^2 \right)
\ee
can be interpreted
as a measure of the quantum contribution to the energy dispersion of the state
$\sigma_{R,t}$.


We shall now illustrate the above results considering states of the form
\be
\sigma_{R,t} =  \alpha |\psi(t)\rangle \langle \psi(t)| + \frac{(1-\alpha)}{d} \mathbb{I}_d,
\ee
where
$0 \le \alpha \le 1$, and $\mathbb{I}_d$ is the $d\times d$ identity matrix (recall that $d$ is the dimension
of $\mathcal{H}_R$).
These states can be regarded as pure states perturbed by white noise. We decompose $|\psi(t)\rangle$ as

\begin{equation}
|\psi(t)\rangle=\sum_{n}c_n e^{-iE_nt/\hbar}|\phi_n\rangle,
\end{equation}
where $\{|\phi_n\rangle\}$ is the set of (orthonormal) eigenstates of $\hat H_R$ with corresponding eigenvalues $E_n$, and the normalization condition $\sum_n |c_n|^2=1$ is satisfied.

Direct calculation gives
\begin{equation}
 \textrm{Tr}\,\sigma_{R,t}^2= \alpha^2 + \frac{1}{d}(1-\alpha^2),
\end{equation}
and
\ben
 \textrm{Tr}\,\sigma^2_{R|M}&=& \sum_E\left[\alpha^2 \sum_{\substack{n\\ (E_n=E)}}|c_n|^2 \sum_{\substack{m\\ (E_m=E)}}|c_m|^2\right]\nonumber\\ &+& 2 \frac{\alpha(1-\alpha)}{d}\underbrace{\sum_E p_E}_{=1} + \frac{(1-\alpha)^2}{d}  \nonumber\\
 &=& \sum_E\left[\alpha^2 p^2_E\right]  + \frac{1}{d}(1-\alpha)^2,
\een
where
\be
p_E=\sum_{\substack{n \\ (E_n=E)}}|c_n|^2. \nonumber
\ee
Using Eq. (34) we thus get
\ben
\Delta S_\textrm{max} &=& \textrm{Tr}\,\sigma_{R,t}^2-\textrm{Tr}\,\sigma^2_{R|M}\nonumber\\
&=& S_L[\alpha\, p_E] + (\alpha^2-1) \nonumber\\
&=& \alpha^2S_L[\{ p_E \} ],
\een
and therefore recover the result (\ref{deltamaxP}) for $\alpha=1$.

On the other hand, one also has
\ben
\textrm{Tr}\left([\hat{H}_R,\sigma_{R,t}]^2\right)&=&\alpha^2\,\textrm{Tr}\left(\big[\hat{H}_R,|\psi\rangle\langle\psi|\big]^2\right)\nonumber\\
&=&-2\alpha^2 \sigma_E^2,
\een
which reduces for $\alpha=1$ to the expression (\ref{trpure}) corresponding to pure states.

 In summary, the entropic indicator $\Delta S$ that detects  entanglement between the system $R$ and the clock $C$ is given, to lowest order in the length $T$ of the interval within which the state of $R+C$ is defined, by a quantity representing the quantum contribution to the energy uncertainty of $R$.


\section{Concluding remarks}

  In quantum mechanics, as in life, it takes two to tango.
  Time evolution requires a composite consisting of at least two parts:
  a system $R$ that evolves, and a system $C$, the clock,
  that keeps track of time. All the properties of the dynamical evolution
  of $R$ can be encoded in the correlations (entanglement)
  exhibited by a stationary quantum state jointly describing
  the complete system $R+C$. In this sense, the origins of dynamics
  and of the flow of time  are, perhaps, the most radical instances of
  the central role played
  by entanglement in the physics of composite quantum systems. These
  considerations constitute the gist of the timeless picture of
  quantum dynamics. According to this viewpoint, there have to be
  quantitative relations connecting the amount of entanglement between
  the clock and the evolving system, on the one hand, with specific features
  of the dynamical evolution of the system, on the other one.
  
  In the present contribution we explore these relations for an
  extension of the PW proposal which, while allowing mixed quantum states
  of the $R+C$ composite, keeps the other PW main assumption, particularly
  that concerning a definite energy of the $R+C$ system equal to zero.
  By recourse to an entanglement indicator for the global state of $R+C$,
  it is possible to elucidate how entanglement relates to the time evolution
   of the system $R$. It turns out that, in our extension of the PW scenario,
   entanglement is indeed
  necessary for $R$ to exhibit evolution. That is, mild forms of quantum correlations,
  such as quantum discord without entanglement, are not enough to give rise
  to time and dynamics. This conclusion follows from an entropic
  sufficient criterium for entanglement satisfied by the state
  (pure or mixed) of $R+C$ whenever the system $R$ exhibits
  dynamical evolution.

  It is a fact of the quantum world that dynamical evolution is always accompanied
  by energy uncertainty. Consistently, the system-clock entanglement is related
  to energy uncertainty as well. Indeed, the aforementioned entanglement indicator
  for  global states of $R+C$ admits an upper bound and an asymptotic limit,
  both expressible in terms of the spread of
  the energy probability distribution associated with the system $R$,
  as measured by an entropic measure evaluated on that distribution.
  The entanglement indicator is also related to the energy dispersion of system $R$,
  in a way reminiscent of a time-energy uncertainty relation.

  Our present developments suggest various possible lines for further inquiry.
  In would be interesting to explore mixed-state formulations of the PW
  timeless approach for systems with a clock having a discrete, finite Hilbert space.
  For this kind of systems, mixed-state timeless scenarios may be amenable
  to experimental implementation, leading to extensions
  of the works reported in \cite{MSVW2015,PRBGBRL2019}. On the theoretical side,
  to include mixed states within the timeless picture may contribute to
  elucidate the way in which the PW approach
  is related to quantum thermodynamics, and to quantum coherence \cite{MS2019}.
  In this regard, it would be interesting to explore more general extensions
  of the PW formulation, allowing for mixed states of the system-clock compound having
  a non-vanishing dispersion of the total energy. These lines
  of inquiry, in turn, may be enriched by including relativistic effects, along the
  lines pioneered in \cite{DR2019,DMR2019}.


\begin{acknowledgements}A.V.H. acknowledges financial support from DGAPA-UNAM through project PAPIIT IN113720. A.P.M acknowledges the Argentinian agencies SeCyT-UNC and CONICET for financial support.
\end{acknowledgements}

\vspace{20pt}


\end{document}